\documentclass[12pt]{ronbun}

\usepackage{amsmath,amssymb,graphicx,color}
\usepackage{cite}
\usepackage{bm}
\usepackage{dcolumn}

\newcommand{\nn}{\nonumber}
\newcommand{\e}{{\rm e}}

\newcommand{\al}{\alpha}

\numberwithin{equation}{section}

\begin{document}

\begin{flushright}
\parbox{4.2cm}
{KEK-TH-1049 \hfill \\
{\tt hep-th/0511072}
 }
\end{flushright}

\vspace*{1.1cm}

\begin{center}
 \Large\bf  Point-Like Graviton Scattering \\ 
in Plane-Wave Matrix Model 
\end{center}
\vspace*{1.0cm}
\centerline{\large  Hyeonjoon Shin$^{\dagger\ast a}$ and Kentaroh
Yoshida$^{\ddagger b}$}
\begin{center}
$^{\dagger}$\emph{CQUeST K209,
Sogang University,
Seoul 121-742, South Korea 
}\\ 
$^{\ast}$\emph{BK 21 Physics Research
    Division and Institute of Basic Science\\ 
    Sungkyunkwan University,
    Suwon 440-746, South Korea 
}\\ 
$^{\ddagger}$\emph{Theory Division, High Energy Accelerator Research 
Organization (KEK),\\
Tsukuba, Ibaraki 305-0801, Japan.} 
\\
\vspace*{1cm}
 $^{a}$hshin@sogang.ac.kr \qquad $^{b}$kyoshida@post.kek.jp
\end{center}

\vspace*{1.0cm}

\centerline{\bf Abstract}
\vspace*{0.5cm} 
In a plane-wave matrix model we discuss a two-body scattering of
gravitons in the $SO(3)$ symmetric space. In this case the graviton
solutions are point-like in contrast to the scattering in the $SO(6)$
symmetric space where spherical membranes are interpreted as
gravitons. We concentrate on a configuration in the 1-2 plane where a
graviton rotates with a constant radius and the other one elliptically
rotates. Then the one-loop effective action is computed by using the
background field method. As the result, we obtain the $1/r^7$-type
interaction potential, which strongly suggests that the scattering in
the matrix model would be
closely related to that in the light-front eleven-dimensional
supergravity. 

\vfill

\thispagestyle{empty}
\setcounter{page}{0}

\newpage 

\section{Introduction and Summary} 

M-theory is considered as the unified theory of superstring
theories. The basic degrees of freedom of string theory and M-theory are
fully encoded in matrix models \cite{BFSS,IKKT,DVV}. The matrix model
approaches lead to non-perturbative formulations of superstring theory
and M-theory. For example, the BFSS matrix model is a supersymmetric
matrix quantum mechanics, which is believed to be a discrete light-cone
quantized M-theory (light-front M-theory). The matrix model also
describes the low-energy dynamics of $N$ D0-branes of type IIA
superstring theory \cite{Witten}. Furthermore it goes to the light-cone
action for the supermembrane in eleven dimensions \cite{dWHN} in the
large $N$ limit. The same type of matrix model as the BFSS matrix model
can be obtained from the supermembrane theory via a matrix regularization
\cite{dWHN}.

\vspace*{0.3cm}

The matrix model on a pp-wave background was proposed by
Berenstein-Maldacena-Nastase (BMN) \cite{BMN}, and it is often called
plane-wave matrix model or BMN matrix model. The pp-wave 
background is given by the following metric and the constant 
four-form field strength \cite{KG}:
\begin{eqnarray}
\label{pp}
ds^2 &=& -2dx^+dx^- -\left(
\sum_{i=1}^3\left(\frac{\mu}{3}\right)^2 (x^i)^2 + \sum_{a=4}^6\left(
\frac{\mu}{6}\right)^2 (x^a)^2
\right)(dx^+)^2 + \sum_{I=1}^9(dx^{I})^2\,, \\
F_{+123} &=& \mu\,. \nn
\end{eqnarray}
This background is maximally supersymmetric and preserves 32
supersymmetries. The action of the matrix model is given
by\footnote{Hereafter we will rescale the gauge field and parameters as
$ A \rightarrow R A\,,~ t \rightarrow t/R\,,~ \mu \rightarrow R \mu$\,.}
\begin{eqnarray}
&& S_{\rm pp} = \int\!\! dt\, 
\mathrm{Tr} \Bigl[ \frac{1}{2R} D_t X^I D_t X^I + \frac{R}{4} ( [
X^I, X^J] )^2 + i \Theta^\dagger D_t \Theta - R \Theta^\dagger \gamma^I
[ \Theta, X^I ]  \nn \\ 
&& \qquad \quad -\frac{1}{2R} \left( \frac{\mu}{3} \right)^2 (X^i)^2
-\frac{1}{2R} \left( \frac{\mu}{6} \right)^2 (X^a)^2 - i \frac{\mu}{3}
\epsilon^{ijk} X^i X^j X^k - i \frac{\mu}{4} \Theta^\dagger \gamma^{123}
\Theta \Bigr]\,, 
\end{eqnarray}
where the indices of the transverse nine-dimensional space are
$I,J=1,\ldots,9$ and $R$ is the radius of the circle compactified along
$x^-$\,. All degrees of freedom are $N\times N$ Hermitian matrices and
the covariant derivative $D_t$ with the gauge field $A$ is defined by
$D_t = \partial_t-i[A,~~]$\,. The plane-wave matrix model can be
obtained from the supermembrane theory on the pp-wave background
\cite{DSR,SY} via the matrix regularization \cite{dWHN}.  In particular,
in the case of the pp-wave, the correspondence of superalgebra
\cite{BSS} between the supermembrane theory and the matrix model,
including brane charges, is established by the works \cite{SY} and
\cite{HS1}. Then an $\mathcal{N}=(4,4)$ type IIA string theory can be
constructed from the supermembrane theory on the pp-wave
\cite{SY4,HS2}. The corresponding matrix string theory on
the pp-wave has been constructed in \cite{SY4} by using the method 
\cite{Sekino} (For other matrix string
theories on pp-waves, see \cite{Bonelli}).

This matrix model may be considered as a deformation of the BFSS matrix
model while it still preserves 32 supersymmetries. The plane-wave matrix
model allows a static 1/2 BPS fuzzy sphere with zero light-cone energy
to exist as a classical solution, since the action of the matrix model
includes the Myers term \cite{Myers}. The structure of the vacua of the
plane-wave matrix model is enriched by the fuzzy sphere. The spectra
around the vacua are now fully clarified \cite{DSR,DSR2,KP,KP2}. The
trivial vacuum $X^I=0$ has also been identified with a single spherical
five-brane vacuum in \cite{TM5}. Except for the static fuzzy sphere,
there are various classical solutions and those are well studied
\cite{Bak,HS1,Park,sol}. Stabilities of the fuzzy sphere are shown in
several papers \cite{DSR,SY3,HSKY}. Thermal stabilities of classical
solutions are also investigated in \cite{Huang,HSKY-thermal,Furuuchi}.

\vspace*{0.3cm}

In our previous papers \cite{HSKY,HSKY-potential}, we have discussed a
two-body scattering of spherical membranes which are considered as giant
gravitons. Then we considered a configuration in a sub-plane in the
$SO(6)$ symmetric space where a spherical membrane (with $p^+=N_1/R$)
rotates with a constant radius $r_1$ and another one (with $p^+=N_2/R$)
elliptically rotates with $r_2 \pm \epsilon$\,. For this setup we have
computed the effective action by using the background field method. The
resulting effective action with respect to $r \equiv r_2 - r_1$
is\footnote{In fact, $r$ should be understood as $|r_2-r_1|$\,, as noted in
\cite{HSKY-potential}.}
\begin{align}
\Gamma_{\rm eff} = \epsilon^4 \int dt \left[
 \frac{35}{2^7 \cdot 3} \frac{N_1 N_2}{r^7}
-\frac{385}{2^{11} \cdot 3^3}
  \big[ 2 (N_1^2 + N_2^2) -1 \big] \frac{N_1 N_2}{r^9}
+  \mathcal{O} \left( \frac{1}{r^{11}} \right) \right]
+ \mathcal{O} ( \epsilon^6)\,. \nn 
\end{align}
This result strongly suggests that the spherical membranes should be
interpreted as spherical gravitons as discussed by Kabat and Taylor
\cite{KT}. Here we should remark that the subleading term is $1/r^9$ and
it is repulsive. In the BFSS case the subleading term is $1/r^{11}$
order and it implies the dipole-dipole interaction. According to the
interpretation, the $1/r^9$ term would imply the dipole-graviton
interaction. This is a new effect intrinsic to the pp-wave background.

\vspace*{0.3cm}

\begin{figure}
 \begin{center}
  \includegraphics[scale=.6]{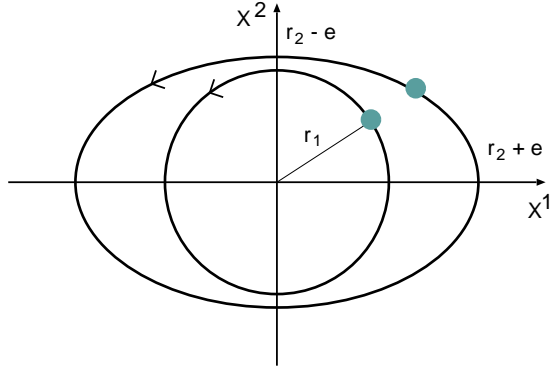} 
 \end{center}
\vspace*{-1cm}
\caption{The configurations of two gravitons.}
\label{gravs:fig} 
\end{figure}

In this paper we will discuss a two-body scattering in the $SO(3)$
symmetric space. Then the configuration for the computation consists of
two {\it point-like} gravitons in contrast to the spherical membrane
cases. The one rotates with a constant radius $r_1$ and the other
elliptically rotates with $r_2\pm\epsilon$\,, as drawn in Fig.\
\ref{gravs:fig}. The resulting effective action is obtained as
\begin{align}
\Gamma_{\rm eff} &=
\epsilon^4\int dt \left[~\frac{35}{24} \frac{1}{r^7}
+ \frac{385}{576} \frac{1}{r^9} + \mathcal{O}\left(\frac{1}{r^{11}}
\right)~\right] + \mathcal{O}(\epsilon^6)\,. \nn
\end{align}
In contrast to the spherical membrane cases, the subleading term becomes 
attractive. 

\vspace*{0.3cm}
The organization of this paper is as follows: In section 2, by using the
background field method around the setup mentioned above, we compute the
functional determinants. Before performing the path integral for the
fluctuations, it is necessary to take care of the time-dependence of the
configuration of classical solutions. In section 3 the functional
determinants are evaluated by expanding them with respect to the
infinitesimal parameter $\epsilon$\,. The evaluation is too complicated,
and so we use the Mathematica \cite{wolfram}. The resulting effective
action gives rise to the $1/r^7$-type potential as the leading
term. Section 4 is devoted to a conclusion and discussions.

\section{Two-Body Interaction of Point-Like Gravitons}

From now on, let us examine the interaction potential between the
point-like gravitons by using the setup proposed in
Fig.\,\ref{gravs:fig}. We will use the background field method as
usual. Then the matrix fields are decomposed into backgrounds and 
fluctuations as follows: 
\begin{eqnarray}
\label{cl+qu}
 X^I = B^I + Y^I\,, \quad \Theta = 0 + \Psi\,. 
\end{eqnarray}
Here $B^I$ are classical backgrounds while $Y^I$ and 
$\Psi$ are quantum fluctuations around them. The fermionic background 
is taken to be zero. 

The background for the configuration in Fig.\,\ref{gravs:fig} is
described by the following $2\times 2$ matrices: 
\begin{eqnarray}
&& B^I = \begin{pmatrix} 
B^I_{(1)}  &  0 \\ 0 & B^I_{(2)} 
\end{pmatrix}
 \qquad (I=1,\ldots,9)\,, \nn \\
&& B^{1}_{(1)} = r_1\cos\left(\frac{\mu}{3}t\right)\,, \qquad 
B^{2}_{(1)} = r_1 \sin\left(\frac{\mu}{3}t\right)\,, \nn \\ 
&& B^{1}_{(2)} = (r_2+\epsilon)\cos\left(\frac{\mu}{3}t\right)\,, \qquad  
B^2_{(2)} = (r_2-\epsilon)\sin\left(\frac{\mu}{3}t\right) \nn \\
&& B^3_{(s)} = B^a_{(s)} = 0 \qquad (s=1,2~;~a=4,\ldots,9)\,. 
\end{eqnarray}
Two gravitons rotating in the 1-2 plane are diagonally embedded.  
Each of the gravitons carries a unit of the light-cone momentum and 
it is represented by a $1\times 1$ matrix. 
One of them rotates with a constant radius $r_1$ 
and the other one rotates elliptically with $r_2\pm\epsilon$\,. 

In order to perform the path integral for the fluctuations, we need to fix
the gauge symmetry. In the matrix model computation, it is convenient
to choose the background field gauge,
\begin{equation}
D_\mu^{\rm bg} A^\mu_{\rm qu} \equiv D_t A + i [ B^I, X^I ] = 0\,.
\label{bg-gauge}
\end{equation}
Then the corresponding gauge-fixing $S_\mathrm{GF}$ and Faddeev-Popov
ghost $S_\mathrm{FP}$ terms are given by
\begin{equation}
S_\mathrm{GF} + S_\mathrm{FP} = \int\!dt \,{\rm Tr} \left( - \frac{1}{2}
(D_\mu^{\rm bg} A^\mu_{\rm qu} )^2 - \bar{C} \partial_{t} D_t C + [B^I,
\bar{C}] [X^I,\,C] \right)\,.  \label{gf-fp}
\end{equation}
Now, by inserting the decomposition of the matrix fields (\ref{cl+qu})
into the matrix model action, we get the gauge fixed plane-wave action
$S$ $(\equiv S_{\rm pp} + S_\mathrm{GF} + S_\mathrm{FP})$ expanded around
the background.  The resulting action is read as 
$S =  S_0 + S_2 + S_3 + S_4$\,,  
where $S_n$ represents the action of order $n$ with respect to the
quantum fluctuations and, for each $n$, its expression is
\begin{align}
S_0 = \int dt \, \mathrm{Tr} \bigg[ \,
&      \frac{1}{2}(\dot{B}^I)^2  
        - \frac{1}{2} \left(\frac{\mu}{3}\right)^2 (B^i)^2 
        - \frac{1}{2} \left(\frac{\mu}{6}\right)^2 (B^a)^2 
        + \frac{1}{4}([B^I,\,B^J])^2
        - i \frac{\mu}{3} \epsilon^{ijk} B^i B^j B^k 
    \bigg] ~,
\notag \\
S_2 = \int dt \, \mathrm{Tr} \bigg[ \,
&       \frac{1}{2} ( \dot{Y}^I)^2 - 2i \dot{B}^I [A, \, Y^I] 
        + \frac{1}{2}([B^I , \, Y^J])^2 
        + [B^I , \, B^J] [Y^I , \, Y^J]
        - i \mu \epsilon^{ijk} B^i Y^j Y^k
\notag \\
&       - \frac{1}{2} \left( \frac{\mu}{3} \right)^2 (Y^i)^2 
        - \frac{1}{2} \left( \frac{\mu}{6} \right)^2 (Y^a)^2 
        + i \Psi^\dagger \dot{\Psi} 
        -  \Psi^\dagger \gamma^I [ \Psi , \, B^I ] 
        -i \frac{\mu}{4} \Psi^\dagger \gamma^{123} \Psi  
\notag \\ 
&       - \frac{1}{2} \dot{A}^2  - \frac{1}{2} ( [B^I , \, A])^2 
        + \dot{\bar{C}} \dot{C} 
        + [B^I , \, \bar{C} ] [ B^I ,\, C] \,
     \bigg] ~,
\notag \\
S_3 = \int dt \, \mathrm{Tr} \bigg[
&       - i\dot{Y}^I [ A , \, Y^I ] - [A , \, B^I] [ A, \, Y^I] 
        + [ B^I , \, Y^J] [Y^I , \, Y^J] 
        +  \Psi^\dagger [A , \, \Psi] 
\notag \\
&       -  \Psi^\dagger \gamma^I [ \Psi , \, Y^I ] 
        - i \frac{\mu}{3} \epsilon^{ijk} Y^i Y^j Y^k
        - i \dot{\bar{C}} [A , \, C] 
        +  [B^I,\, \bar{C} ] [Y^I,\,C]  \,
     \bigg] ~,
\notag \\
S_4 = \int dt \, \mathrm{Tr} \bigg[
&       - \frac{1}{2} ([A,\,Y^I])^2 + \frac{1}{4} ([Y^I,\,Y^J])^2 
     \bigg] ~.
\label{bgaction} 
\end{align}
Here the action of the first order becomes zero by using the equations
of motion. 

For the justification of one-loop computation or the semi-classical
analysis, it should be made clear that $S_3$ and $S_4$ can be regarded
as perturbations.  For this purpose, following \cite{DSR}, we rescale
the fluctuations and parameters as
\begin{eqnarray}
&& A   \rightarrow \mu^{-1/2} A\,, \quad 
Y^I \rightarrow \mu^{-1/2} Y^I\,, \quad 
C  \rightarrow \mu^{-1/2} C\,, \quad 
\bar{C} \rightarrow \mu^{-1/2} \bar{C}\,, \nn \\
&& r_{1,2} \rightarrow \mu r_{1,2}\,, \quad \epsilon \rightarrow
 \mu\epsilon\,, \quad t \rightarrow \mu^{-1} t\,. 
\label{rescale}
\end{eqnarray}
Under this rescaling, the action $S$ becomes
\begin{align}
S =  S_2 + \mu^{-3/2} S_3 + \mu^{-3} S_4\,,
\label{ssss}
\end{align}
where the parameter $\mu$ in $S_2$, $S_3$ and $S_4$ has been replaced by
1 and so those do not have $\mu$ dependence.  Now it is obvious that, in
the large $\mu$ limit, $S_3$ and $S_4$ can be treated as perturbations
and the one-loop computation gives the sensible result. Note that the
analysis in the $S_2$ part is exact in the $\mu\rightarrow \infty$
limit. 

Based on the structure of the classical background, we now take the 
quantum fluctuations in the $2\times 2$ off-diagonal matrices:  
\begin{eqnarray}
&& A = \begin{pmatrix}
0 & \Phi^0 \\ \Phi^{0\dagger} & 0 
\end{pmatrix}
\,,  \quad Y^I = \begin{pmatrix} 
0 & \Phi^I \\ \Phi^{I\dagger} & 0 
\end{pmatrix}
\,, \quad \Psi = \begin{pmatrix}
0 & \chi \\ 
\chi^{\dagger} & 0 \\
\end{pmatrix}
\,, \nn \\ && 
C = \begin{pmatrix}
0 & C \\ C^{\dagger} & 0 
\end{pmatrix}
\,, \quad \bar{C} = \begin{pmatrix}
0 & \bar{C} \\ \bar{C}^{\dagger} & 0  
\end{pmatrix}
\,.
\end{eqnarray}
Here we are interested in the interaction between the gravitons 
and so we set the diagonal components to zero. 
It is an easy task to show the quantum stability of each of the
gravitons by following the method in \cite{HSKY,HSKY-potential}. 

It is convenient to introduce the following quantities:
 \begin{eqnarray}
r\equiv r_2-r_1\,, \qquad g(t) = \epsilon^2 + 2\epsilon r
\cos\left(\frac{2}{3}t\right)\,, \quad 
G_m \equiv \frac{1}{\partial_t^2 + r^2 + m^2}\,. 
\end{eqnarray}
Here $G_m$ is a propagator for a mass $m$\,. 
By using them we can express the functional determinants after the path
integral in simpler forms.  We will perform the path integral below for
each of the parts, bosons, ghosts and fermions.

\subsection{Boson Fluctuation} 

Let us first consider the bosonic parts. The Lagrangian
$L_{\rm B}$ is composed of two parts: 
\begin{eqnarray} 
 L_{\rm B} &=& L_{SO(3)} + L_{SO(6)}\,, \\ 
 L_{SO(3)} &=& - |\dot{\Phi}^0|^2 + \left(
r^2+g(t)\right)|\Phi^0|^2  + |\dot{\Phi}^i|^2 - (r^2 + g(t))|\Phi^i|^2 
- \frac{1}{3^2}|\Phi^i|^2   \\ 
&& + \frac{2}{3}i(r+\epsilon)\sin\left(\frac{t}{3}\right)
(\Phi^{0\dagger}\Phi^1-\Phi^{1\dagger}\Phi^0) 
 - \frac{2}{3}i(r-\epsilon)\cos\left(\frac{t}{3}\right)
\left(\Phi^{0\dagger}\Phi^2-\Phi^{2\dagger}\Phi^0\right) \nn \\
&& -i(r+\epsilon)\cos\left(\frac{t}{3}\right)(\Phi^{2\dagger}\Phi^3
-\Phi^{3\dagger}\Phi^2)
-i(r-\epsilon)\sin\left(\frac{t}{3}\right)(\Phi^{3\dagger}\Phi^1
-\Phi^{1\dagger}\Phi^3)\,, \nn \\ 
L_{SO(6)} &=&  |\dot{\Phi}^a|^2 - (r^2 + g(t))|\Phi^a|^2 
- \frac{1}{6^2}|\Phi^a|^2\,. 
\end{eqnarray}
Here the gauge field is included in $L_{SO(3)}$\,. 
The next task is to evaluate each of the parts.

\subsubsection*{SO(3) part} 

Now we shall consider the $SO(3)$ part. In the Lagrangian for the
$SO(3)$ part, the four variables $\Phi^0$\,, $\Phi^i$~$(i=1,2,3)$ are
contained. The analysis of this part is complicated since these
are coupled. In order to carry out the path integral, it is convenient
to decouple the variables as much as possible. 
For this purpose we first take the coordinate transformation 
\begin{eqnarray}
&& \Phi^1 \equiv \cos\left(\frac{t}{3}\right)\Phi^r 
- \sin\left(\frac{t}{3}\right)\Phi^{\theta}\,, \quad 
\Phi^2 \equiv \sin\left(\frac{t}{3}\right)\Phi^r 
+ \cos\left(\frac{t}{3}\right)\Phi^{\theta}\,, 
\end{eqnarray}
and introduce the new variables $\Phi^r$ and $\Phi^{\theta}$ instead of
$\Phi^1$ and $\Phi^2$\,. 
The Lagrangian after the transformation is rewritten as 
\begin{eqnarray}
L_{SO(3)} &=& -|\dot{\Phi}^0|^2 + (r^2+g(t))|\Phi^0|^2 + 
|\dot{\Phi}^r|^2 - (r^2 + g(t))|\Phi^r|^2 + |\dot{\Phi}^{\theta}|^2 
- (r^2 + g(t))|\Phi^{\theta}|^2 \nn \\ 
&& + |\dot{\Phi}^3|^2 - (r^2+g(t) + (1/3^2))|\Phi^3|^2 + 
\frac{2}{3}(\Phi^{r\dagger}\dot{\Phi}^{\theta} 
- \Phi^{\theta\dagger}\dot{\Phi}^r) \nn \\ 
&& +
 \frac{2}{3}i\epsilon\sin(2t/3)(\Phi^{0\dagger}\Phi^r - \Phi^{r\dagger}\Phi^0) 
- \frac{2}{3}i(r - \epsilon\cos(2t/3))(\Phi^{0\dagger}\Phi^{\theta}
- \Phi^{\theta\dagger}\Phi^0) \nn \\ 
&&  +
i\epsilon\sin(2t/3)(\Phi^{3\dagger}\Phi^r - \Phi^{r\dagger}\Phi^3) 
+ i(r + \epsilon\cos(2t/3))(\Phi^{3\dagger}\Phi^{\theta}
- \Phi^{\theta\dagger}\Phi^3)\,. 
\end{eqnarray} 
Taking the shift of $\Phi^0$ and $\Phi^3$ defined by 
\begin{eqnarray}
&& \Phi^0 \equiv \Phi^0{}' - \frac{2}{3}i(G_0^{-1} + g(t))^{-1}\left[
\epsilon \sin\left(\frac{2}{3}t\right)\Phi^r - (r-\epsilon\cos\left(\frac{2}{3}t\right))\Phi^{\theta}
\right]\,, \nn \\ 
&& \Phi^3 \equiv \Phi^3{}' + i(G_{1/3}^{-1} + g(t))^{-1}\left[
\epsilon \sin\left(\frac{2}{3}t\right)\Phi^r + (r+\epsilon\cos\left(\frac{2}{3}t\right))\Phi^{\theta}
\right]\,, \nn
\end{eqnarray}
we obtain the following Lagrangian:
\begin{eqnarray}
&& \hspace*{-1cm} 
L_{SO(3)} = \Phi^{0}{}'{}^{\dagger}(G_0^{-1} +g(t))\Phi^{0}{}' 
- \Phi^3{}'{}^{\dagger}(G_{1/3}^{-1}+g(t))\Phi^3{}' - \Phi^{r\dagger}(G_0^{-1}+g(t))\Phi^r 
\nn \\ 
\qquad 
&& - \Phi^{\theta\dagger}(G_0^{-1} +g(t))\Phi^{\theta} 
+ \frac{2}{3}\Phi^{r\dagger}\dot{\Phi}^{\theta} 
-\frac{2}{3} \Phi^{\theta\dagger}\dot{\Phi}^r  \\
\qquad 
&& - \Phi^{r\dagger}\left\{\epsilon^2\sin\left(2t/3\right) 
\left[
\frac{4}{9}(G_0^{-1} + g(t))^{-1} - (G_{1/3}^{-1} + g(t))^{-1}\right]
\sin\left(2t/3\right)\right\}\Phi^r \nn \\
\qquad 
&& - \Phi^{\theta\dagger}\left\{
\left(r-\epsilon\cos\left(2t/3\right)\right) 
\left[\frac{4}{9}(G_0^{-1} + g(t))^{-1} - (G_{1/3}^{-1}+g(t))^{-1} \right]
\left(r-\epsilon\cos\left(2t/3\right)\right) 
\right\}\Phi^{\theta} \nn \\ 
\qquad 
&& + \Phi^{\theta\dagger}\left\{
\left(r-\epsilon\cos\left(2t/3\right)\right) \left[
\frac{4}{9}(G_0^{-1} + g(t))^{-1} + (G_{1/3}^{-1}+g(t))^{-1}\right]
\epsilon\sin(2t/3) \right\}\Phi^r \nn \\ 
\qquad 
&& + \Phi^{r\dagger}\left\{\epsilon\sin\left(2t/3\right) 
\left[ \frac{4}{9}(G_0^{-1} + g(t))^{-1} + (G_{1/3}^{-1} +g(t))^{-1} \right] 
\left(r-\epsilon\cos\left(2t/3\right)\right) 
\right\}\Phi^{\theta}\,. \nn 
\end{eqnarray} 
Note that $\Phi^0$ and $\Phi^3$ are decoupled from $\Phi^r$ and
$\Phi^{\theta}$ at this stage, but $\Phi^r$ and $\Phi^{\theta}$ 
are still coupled. We can however perform the path integral for $\Phi^r$ and
$\Phi^{\theta}$ by using the formula 
\begin{eqnarray}
\begin{pmatrix}
A & B \\ C & D 
\end{pmatrix}
= \begin{pmatrix}
A & 0 \\ C & 1 
\end{pmatrix}
\begin{pmatrix}
1 & A^{-1}B \\ 0 & D-CA^{-1}B 
\end{pmatrix}
\,. \label{formula}
\end{eqnarray}
The resulting effective action for the $SO(3)$ part is given by 
\begin{eqnarray} 
&& \hspace*{-0.75cm} \e^{i\Gamma_{SO(3)}} = \left[
\det(G_0^{-1} + g(t))\cdot\det(G_{1/3}^{-1}+g(t))\cdot 
\det A \cdot\det(D-CA^{-1}B)\right]^{-1}  
\,,  \\ 
\quad && A = G_0^{-1} + g(t) + \epsilon^2\sin\left(2t/3\right) 
\left[
\frac{4}{9}\,\frac{1}{G_0^{-1} + g(t)} - \frac{1}{G_{1/3}^{-1} + g(t)}\right]
\sin\left(2t/3\right)\,, \nn \\ 
\quad && D = G_0^{-1} + g(t) 
+ \frac{4}{9} \left(r-\epsilon\cos\left(2t/3\right)\right) 
\frac{1}{G_0^{-1} + g(t)} 
\left(r-\epsilon\cos\left(2t/3\right)\right) \nn \\
&& \qquad\qquad  - \left(r+\epsilon\cos\left(2t/3\right)\right) 
\frac{1}{G_{1/3}^{-1}+g(t)}\left(r+\epsilon\cos\left(2t/3\right)\right) 
\,, \nn \\ 
\quad 
&& B = -\frac{2}{3}\partial_t - \frac{4}{9}\epsilon\sin\left(2t/3\right) 
\frac{1}{G_0^{-1} + g(t)}
\left(r-\epsilon\cos\left(2t/3\right)\right) \nn \\ 
&& \qquad\qquad  
- \epsilon\sin\left(2t/3\right)\frac{1}{G_{1/3}^{-1} +g(t)} 
\left(r+\epsilon\cos\left(2t/3\right)\right)\,, 
\nn \\ 
\quad && C = \frac{2}{3}\partial_t - \frac{4}{9}
\left(r-\epsilon\cos\left(2t/3\right)\right)
\frac{1}{G_0^{-1} + g(t)}\epsilon\sin(2t/3) \nn \\ 
&& \qquad \qquad 
- \left(r+\epsilon\cos\left(2t/3\right)\right)
\frac{1}{G_{1/3}^{-1}+g(t)}\epsilon\sin(2t/3)\,. \nn 
\end{eqnarray}
Then we will examine the $SO(6)$ part. 

\subsubsection*{SO(6) part}

It is straightforward to perform the path integral for the $SO(6)$ part. 
The result is 
\begin{eqnarray}
\e^{\Gamma_{SO(6)}} = \e^{i\Gamma_{SO(6)}^{(0)}}
(\det\left(1+G_{1/6}\,g(t)\right))^{-6}\,,
\end{eqnarray}
where the $\epsilon$-independent part is written as 
\begin{eqnarray}
\e^{i\Gamma^{(0)}_{SO(6)}} = (\det G_{1/6}^{-1})^{-6}\,. 
\end{eqnarray}

\subsection{Ghost Fluctuation} 

Next we shall consider the ghost part. 
The Lagrangian for the ghost part is given by 
\begin{eqnarray}
L_{\rm G} = 
\dot{\bar{C}}\dot{C}^{\dagger} + \dot{\bar{C}}^{\dagger}\dot{C} 
- (r^2 + g(t))(\bar{C}C^{\dagger} + \bar{C}^{\dagger}C)\,. 
\label{gho} 
\end{eqnarray}
The path integral for (\ref{gho}) is immediately evaluated as 
\begin{eqnarray}
[\det\left(G_0^{-1} + g(t)\right)]^2\,. 
\end{eqnarray}

\subsection{Fermion Fluctuation} 

Finally, let us discuss the fermionic part. 
The Lagrangian for the fermion fluctuations is given by 
\begin{eqnarray}
L_{\rm F} &=& 2\Bigl[i\chi^{\dagger}\dot{\chi} -r\chi^{\dagger}\left(
\gamma^1\cos\left(\frac{t}{3}\right) + \gamma^2 \sin\left(\frac{t}{3}\right)
\right)\chi \nn \\ 
&& - \epsilon\chi^{\dagger}\left(\gamma^1\cos\left(\frac{t}{3}\right) - \gamma^2\sin\left(\frac{t}{3}\right)\right)\chi 
- \frac{i}{4}\chi^{\dagger}\gamma^{123}\chi \Bigr]\,. 
\end{eqnarray}
Then we decompose the spinor $\chi$ into two components as follows: 
\begin{eqnarray}
\chi = \dbinom{\chi_{A\al}}{\hat{\chi}_{\al}^A}\,, \qquad \hat{\chi}_{\al}^A 
\equiv \epsilon_{\al\beta}\hat{\chi}^{A\beta}\,.  
\end{eqnarray} 
According to this decomposition, the $SO(9)$ gamma matrices should also
be decomposed. In our analysis only $\gamma^1$\,, $\gamma^2$ and
$\gamma^3$ are necessary and hence we write down only them here,  
\begin{eqnarray} 
\gamma^i = \begin{pmatrix} 
-\sigma^i \times 1 & 0 \\ 0 & \sigma^i \times 1 \\ 
\end{pmatrix}
\qquad (i=1,2,3)\,. 
\end{eqnarray} 
For the detail of the decomposition of the gamma matrices, see
\cite{DSR,HSKY}.  The Lagrangian after the decomposition is written as
\begin{eqnarray}
L_{\rm F} &=& i\chi^{\dagger A\al}\dot{\chi}_{A\al} 
- \frac{1}{4}\chi^{\dagger A\al}\chi_{A\al} + i\hat{\chi}^{\dagger\al}_A\dot{\hat{\chi}}_{\al}^A  
+ \frac{1}{4}\hat{\chi}^{\dagger\al}_A\hat{\chi}_{\al}^A  \\ 
&& + \chi^{\dagger A\al}\left[ 
(r+\epsilon) \cos(t/3)\cdot \sigma^1 + (r-\epsilon) \sin(t/3)\cdot 
\sigma^2\, 
\right]_{\al}^{~\beta}\chi_{A\beta} \nn \\
&& - \hat{\chi}^{\dagger\al}_A\left[ 
(r+\epsilon) \cos(t/3)\cdot \sigma^1 + (r-\epsilon) \sin(t/3)\cdot 
\sigma^2\, 
\right]_{\al}^{~\beta}\hat{\chi}_{\beta}^A\,. \nn 
\end{eqnarray}
Here it is convenient to introduce new spinors $\chi'$ and
$\tilde{\chi}$ defined by
\begin{eqnarray}
\chi_{A\al} = (\e^{-i\sigma^3 \frac{t}{6}})_{\al}^{~\beta}\,
\chi'_{A\beta}\,, \qquad 
\hat{\chi} = (\e^{-i\sigma^3 \frac{t}{6}})_{\al}^{~\beta}\,\tilde{\chi}_{\beta}^A\,. 
\end{eqnarray}
By using the formulae:
\begin{eqnarray} 
&& \e^{i\sigma^3\frac{t}{6}}\left[\sigma^1 \cos(t/3) + \sigma^2 \sin(t/3) 
\right] \e^{-i\sigma^3\frac{t}{6}} = \sigma^1\,, \nn \\
&& \e^{i\sigma^3\frac{t}{6}}\left[\sigma^1 \cos(t/3) - \sigma^2 \sin(t/3) 
\right] \e^{-i\sigma^3\frac{t}{6}} = \sigma^1 \cos(2t/3) 
- \sigma^2 \sin(2t/3)\,,  
\end{eqnarray}
we can rewrite the Lagrangian as 
\begin{eqnarray}
L_{\rm F} &=& i\chi'{}^{\dagger A\al}\dot{\chi}'_{A\al} - \frac{1}{4}\chi'{}^{\dagger A\al}\chi'_{A\al} + \frac{1}{6}\chi'{\dagger A\al}(\sigma^3)_{\al}^{~\beta}\chi'_{A\beta} 
+ i\tilde{\chi}^{\dagger\al}_A\dot{\tilde{\chi}}_{\al}^A + \frac{1}{4}\tilde{\chi}^{\dagger\al}_A\tilde{\chi}_{\al}^A + \frac{1}{6}\tilde{\chi}^{\dagger A\al}
(\sigma^3)_{\al}^{~\beta}\tilde{\chi}_{A\beta} \nn \\ 
&& + \chi'{}^{\dagger A\al}\left[
\sigma^1 (r + \epsilon\cos(2t/3)) - \sigma^2 \epsilon\sin(2t/3)
\right]_{\al}^{~\beta} \chi'_{A\beta} \nn \\ 
&& - \tilde{\chi}^{\dagger \al}_A \left[
\sigma^1 (r + \epsilon\cos(2t/3)) - \sigma^2 \epsilon\sin(2t/3)
\right]_{\al}^{~\beta} \tilde{\chi}^A_{\beta}\,. \nn
\end{eqnarray}
Then we express the two components 
of the spinors $\chi'$ and $\tilde{\chi}$ as 
\begin{eqnarray}
\chi' = (\pi,\eta)\,, \qquad \tilde{\chi} =
 (\tilde{\pi},\tilde{\eta})\,.  
\end{eqnarray}
When the Lagrangian is described in terms of $\pi,~\eta,~\tilde{\pi},$
and $\tilde{\eta}$\,, it is decomposed into two parts: 
$(\pi,\eta)$-system and $(\tilde{\pi},\tilde{\eta})$-system. 
The Lagrangian for each of these system are given by  
\begin{eqnarray}
L_{\rm F} &=& L_{\pi,\eta} + L_{\tilde{\pi},\tilde{\eta}}\,, \\ 
L_{\pi,\eta} &=& i\pi^{\dagger}\dot{\pi} -\frac{1}{12}\pi^{\dagger}\pi 
+ i\eta^{\dagger}\dot{\eta}-\frac{5}{12}\eta^{\dagger}\eta \nn \\
&& + (r+\epsilon\e^{\frac{2}{3}it})\pi^{\dagger}\eta +
 (r+\epsilon\e^{-\frac{2}{3}it})\eta^{\dagger}\pi\,,  \\ 
L_{\tilde{\pi},\tilde{\eta}} &=& i\tilde{\pi}^{\dagger}\dot{\tilde{\pi}}
 + 
\frac{5}{12}\tilde{\pi}^{\dagger}\tilde{\pi} +
i\tilde{\eta}^{\dagger}\dot{\tilde{\eta}} +
\frac{1}{12}\tilde{\eta}^{\dagger}\tilde{\eta} \nn \\
&& - (r+\epsilon\e^{\frac{2}{3}it})\tilde{\pi}^{\dagger}\tilde{\eta} 
- (r+\epsilon\e^{-\frac{2}{3}it}) \tilde{\eta}^{\dagger}\tilde{\pi}\,. 
\end{eqnarray}
By using the formula (\ref{formula})\,, we can perform the path
integral for $\pi$\,, $\eta$\,, $\tilde{\pi}$ and $\tilde{\eta}$\,.   
The effective action is given by  
\begin{eqnarray}
&& \hspace*{-1cm}
\e^{i\Gamma_{\rm F}} = \det\left[(i\partial_t - \frac{1}{12})(i\partial_t -
     \frac{5}{12})-r^2\right]^4
\det\left[1-\frac{1}{(i\partial_t-\frac{1}{12})
(i\partial_t-\frac{5}{12})-r^2}E\right]^4 \times \nn \\ 
&& \quad \times \det\left[(i\partial_t + \frac{1}{12})(i\partial_t + 
     \frac{5}{12})-r^2\right]^4
\det\left[1-\frac{1}{(i\partial_t+\frac{1}{12})
(i\partial_t+\frac{5}{12})-r^2}\tilde{E}\right]^4\,,  
\end{eqnarray}
where $E$ and $\tilde{E}$ are defined by, respectively, 
\begin{eqnarray}
&& E =
 r\epsilon\e^{\frac{2}{3}it}+(i\partial_t-\frac{1}{12})\epsilon\e^{-\frac{2}{3}it}\frac{1}{i\partial_t
 - \frac{1}{12}}(r+\epsilon\e^{\frac{2}{3}it})\,, \nn \\
&& \tilde{E} =
 r\epsilon\e^{\frac{2}{3}it}+(i\partial_t +\frac{5}{12})\epsilon\e^{-\frac{2}{3}it}\frac{1}{i\partial_t
 + \frac{5}{12}}(r+\epsilon\e^{\frac{2}{3}it})\,. 
\end{eqnarray}

Now we have finished the path integration for the fluctuations. 
The remaining task is to evaluate the functional determinant. 
This will be discussed in the next section.

\section{Effective Action}

From now on we evaluate the determinant factors obtained in the
previous section. In the evaluation we use the formula,
\[
 \det(1+\epsilon g) = \exp\left(\epsilon {\rm
 tr}g-\frac{1}{2}\epsilon^2{\rm tr}g^2 + \cdots\right)\,,
\]
and therefore the resulting effective action is expressed as an 
expansion in terms of $\epsilon$\,, 
\begin{eqnarray}
\Gamma_{\rm eff} = \Gamma^{(0)}_{\rm eff} + \epsilon^2\Gamma^{(2)}_{\rm
 eff} + \epsilon^4 \Gamma^{(4)}_{\rm eff} + \mathcal{O}(\epsilon^6)\,,
\end{eqnarray} 
where the terms of 
order $\epsilon^n$ with odd $n$ are absent in our computation in accordance
with the logic of our previous work \cite{HSKY-potential}.

Before going to the analysis of the $\epsilon$-dependent part, 
let us consider the $\epsilon$-independent part 
$\Gamma^{(0)}_{\rm eff}$ and show the one-loop flatness:  
\[
 \Gamma^{(0)}_{\rm eff} = 0\,. 
\]

\subsection{One-Loop Flatness}

For the $SO(3)$ part, the effective action is  
\begin{eqnarray}
\e^{i\Gamma_{SO(3)}} = \left[\det(G_0^{-1})\cdot\det(G_{1/3}^{-1})\cdot 
\det A \cdot \det(D-CA^{-1}B)\right]^{-1}\,,
\end{eqnarray}
where the components $A,B,C$ and $D$ are given by 
\begin{eqnarray}
A = G_0^{-1}\,, \quad B = -\frac{2}{3}\partial_t\,, \quad
C=\frac{2}{3}\partial_t \quad D = G_0^{-1} + \frac{4}{9}r^2G_0 -r^2
G_{1/3}\,.
\end{eqnarray}
Then we obtain that 
\begin{eqnarray}
D-CA^{-1}B = \partial_t^2 + r^2 + \frac{4}{9} - \frac{r^2}{\partial_t^2+r^2+\frac{1}{9}}\,. 
\end{eqnarray}
Hence we can rewrite a part of the determinant factors as follows:  
\begin{eqnarray}
&& \det(G_{1/3}^{-1})\cdot \det(D-CA^{-1}B) =  
\det\left[\left(\partial_t^2+r^2+\frac{4}{9}\right)
\left(\partial_t^2+r^2+\frac{1}{9}\right)-r^2 
\right]\nn \\ 
&=& \det\left[\partial_t^2 + \left(\partial_t^2 +r^2 -\frac{2}{9}\right)^2
\right] = \det\left[
\left(\partial_t^2 + i\partial_t +r^2 - \frac{2}{9}\right)
\left(\partial_t^2 -i\partial_t +r^2 - \frac{2}{9}\right)
\right]\,. \nn 
\end{eqnarray} 
By using the formula 
\begin{eqnarray}
\int^{\infty}_{-\infty}\frac{dk}{2\pi}\,\ln(-k^2+2pk+m^2-i\epsilon) 
= i\sqrt{m^2+p^2}\,,
\end{eqnarray}
the contribution to the effective action is evaluated as
\begin{eqnarray}
2\sqrt{r^2+\frac{1}{36}}\,. \nn
\end{eqnarray}
This includes the only contribution from the physical degrees of freedom
and the unphysical mode related to the gauge field is surely canceled
out with the ghost contribution.

Turning to the $SO(6)$ part, we see that the contribution from the $SO(6)$
part is given by 
\begin{eqnarray}
6\sqrt{r^2 +\frac{1}{36}}\,.\nn
\end{eqnarray}
Hence the total bosonic contribution is 
\begin{eqnarray}
\Gamma^{(0)}_{\rm B} = 8\sqrt{r^2 +\frac{1}{36}}\,.
\end{eqnarray}

Finally, let us examine the fermionic contribution. 
The effective action is given by 
\begin{eqnarray}
\e^{i\Gamma_{\rm F}} = \det\left[\left(i\partial_t - \frac{1}{12}\right) 
\left(i\partial_t - \frac{5}{12}\right) - r^2 \right]^4 
\cdot \det\left[\left(i\partial_t + \frac{1}{12}\right) 
\left(i\partial_t + \frac{5}{12}\right) - r^2 \right]^4 
\end{eqnarray}
Here, noting that 
\begin{eqnarray}
\left(i\partial_t - \frac{5}{12}\right) - r^2 
= - \left(\partial_t^2 +\frac{1}{2}i\partial_t - \frac{5}{144}\right)\,,\nn
\end{eqnarray}
the total fermionic contribution is evaluated as 
\begin{eqnarray}
\Gamma^{(0)}_{\rm F} = -8\sqrt{r^2 + \frac{1}{36}}\,. 
\end{eqnarray}
Therefore the total contribution from the $\epsilon$-dependent parts  
becomes zero:
\begin{eqnarray}
\Gamma^{(0)}_{\rm eff} &=& \Gamma^{(0)}_{\rm B} + \Gamma^{(0)}_{\rm F}
 \nn \\  
&=& 8\sqrt{r^2 + \frac{1}{36}} -8 \sqrt{r^2 + \frac{1}{36}} = 0\,.  \nn
\end{eqnarray}
Thus, the one-loop flatness:  
\begin{align}
\Gamma^{(0)}_{\rm eff} = 0\,, 
\end{align}
has been shown in our setup.

\subsection{Evaluation of $\epsilon$-Dependent Part}

The evaluation of the $\epsilon$-dependent parts is quite complicated 
and hence we need to use the Mathematica \cite{wolfram}. 
Here we shall show only the results after evaluating the functional
traces.  We note that, due to the enormous complexity of computation,
the results are obtained in the large-distance expansion, $r \gg 1$.

First of all, at $\epsilon^2$ order, we obtain 
\begin{align}
\Gamma^{(2)}_{\rm B} 
&= - \Gamma^{(2)}_{\rm F}  \notag \\
&= - \int dt \left( \frac{2}{r} + \frac{17}{2^2 \cdot 3^2} \frac{1}{r^3}
+ \frac{61}{2^6 \cdot 3^3} \frac{1}{r^5} 
+ \frac{1129}{2^9 \cdot 3^6} \frac{1}{r^7}
+ \frac{26891}{2^{14} \cdot 3^8} \frac{1}{r^9} \right)
+ \mathcal{O} (r^{-11})\,. 
\end{align}
We can see the cancellation between contributions of bosons and
fermions up to  the $1/r^9$ order. 
It is, however, possible to show that the cancellation is exact from the 
numerical analysis. Hence we have shown that the effective action with 
$\epsilon^2$ order should vanish: 
\begin{align}
\Gamma^{(2)}_{\rm eff}=0\,. 
\end{align}

Now let us see the effective action at $\epsilon^4$ order. 
The contribution from the physical modes in the $SO(3)$ part is given by 
\begin{align}
\widehat{\Gamma}^{(4)}_{SO(3)} =
\int dt \left( - \frac{1}{2^5} \frac{1}{r^3}
- \frac{883}{2^8 \cdot 3} \frac{1}{r^5}
+ \frac{11\cdot 19 \cdot 443}{2^{12} \cdot 3^3} \frac{1}{r^7}
+ \frac{11 \cdot 29 \cdot 28793}{2^{15} \cdot 3^6} \frac{1}{r^9}
\right) + \mathcal{O} (r^{-11})\,, 
\end{align}
and, for the $SO(6)$ part, the contribution is  
\begin{align}
\Gamma^{(4)}_{SO(6)} =
\int dt \left( - \frac{3}{2^5} \frac{1}{r^3}
+ \frac{77}{2^8} \frac{1}{r^5}
+ \frac{19 \cdot 131}{2^{12} \cdot 3} \frac{1}{r^7}
+ \frac{13^2 \cdot 4271}{2^{15} \cdot 3^5} \frac{1}{r^9}
\right) + \mathcal{O} (r^{-11})\,. 
\end{align} 
Hence the total contribution of the bosonic parts is   
\begin{align}
\Gamma^{(4)}_{\rm B}
&= \widehat{\Gamma}^{(4)}_{SO(3)} + \Gamma^{(4)}_{SO(6)}
\notag \\
&=
\int dt \left( - \frac{1}{2^3} \frac{1}{r^3}
- \frac{163}{2^6 \cdot 3} \frac{1}{r^5}
+ \frac{17 \cdot 19 \cdot 89}{2^{10} \cdot 3^3} \frac{1}{r^7}
+ \frac{131 \cdot 21661}{2^{13} \cdot 3^6} \frac{1}{r^9}
\right) + \mathcal{O} (r^{-11})\,. 
\end{align}
The contribution of the fermions is totally 
represented by 
\begin{align}
\Gamma^{(4)}_{\rm F} =
\int dt \left( \frac{1}{2^3} \frac{1}{r^3}
+ \frac{163}{2^6 \cdot 3} \frac{1}{r^5}
+ \frac{71 \cdot 163}{2^{10} \cdot 3^3} \frac{1}{r^7}
+ \frac{773 \cdot 1493}{2^{13} \cdot 3^6} \frac{1}{r^9}
\right) + \mathcal{O} (r^{-11})\,. 
\end{align}
Thus the net effective action at $\epsilon^4$ order is given by 
\begin{align}
\Gamma^{(4)}_{\rm eff} &= \Gamma^{(4)}_{\rm B} + \Gamma^{(4)}_{\rm F} 
\notag \\
&=
\int dt \left(\,\frac{35}{24} \frac{1}{r^7}
+ \frac{385}{576} \frac{1}{r^9}\,\right) + \mathcal{O} (r^{-11})\,.
\end{align}
The contributions of the parts with $1/r$\,, $1/r^3$ and $1/r^5$ are
exactly canceled out. These cancellations would be basically due to
the supersymmetries. And the resulting effective potential is
$1/r^7$-type as in the case of the BFSS matrix model.  We should note
that the above expression is written in the Minkowski formulation and so
the leading term of the potential is attractive. Then we should note
that the subleading term of order $1/r^9$ exists and it is also
attractive. Firstly, the term of $1/r^9$-type does not appear in the
BFSS case where the subleading term is $1/r^{11}$ and this corresponds
to the dipole-dipole interaction\,. The presence of the $1/r^9$ term
implies the existence of the dipole-graviton interaction or the
interaction between single poles. The appearance of this term is a new
effect intrinsic to the pp-wave case. Furthermore, we should note that
the subleading term is attractive while the subleading term in the
spherical membrane cases is repulsive. Since the transverse $SO(9)$
symmetry is broken due to the effect of non-vanishing curvature of the
pp-wave background, it is not a mystery to obtain the different graviton
potential in each of the $SO(3)$ and $SO(6)$ symmetric spaces.  It is, 
however, still interesting to see the apparent difference between the graviton
interactions in the $SO(3)$ and the $SO(6)$ symmetric spaces.

\section{Conclusion and Discussion}

We have computed the two-body interaction potential between the
point-like graviton solutions in a sub-plane in the $SO(3)$ symmetric
space by considering the configuration drawn in Fig.\ref{gravs:fig}.
The leading term of the potential is $1/r^7$ and thus strongly suggests
that our result should be closely related to the scattering in the
light-front eleven-dimensional supergravity. We expect that this
potential should be realized from the computation in the supergravity
side by using the spectrum of the linearized supergravity around the
pp-wave background \cite{Kimura}. In this direction the work \cite{LMW}
would be helpful.  


So far we have computed the interaction of the spherical membrane fuzzy
spheres (giant gravitons) in the $SO(6)$ space and the scattering of the
point-like gravitons in the $SO(3)$ symmetric space.  It is interesting
to consider the interaction between a point-like graviton and
a spherical membrane graviton. We hope that the result will be reported in
the near future as another publication \cite{future}. 

Our analysis and results will be an important clue to study some 
features of M-theory on the pp-wave background and to shed light on the
substance of M-theory.

\section*{Acknowledgments} 

One of us, K.~Y., would like to thank Hiroyuki Fuji, Suguru Dobashi,
Naoyuki Kawahara, Jun Nishimura and Xinkai Wu for useful discussions. 
K.~Y. is also grateful to the Third Simons Workshop in Mathematics and
Physics at YITP and the Sapporo Autumn School 2005 at Hokkaido
University for their hospitality.  The work of H.~S. was supported by
grant No. R01-2004-000-10651-0 from the Basic Research Program of the
Korea Science and Engineering Foundation (KOSEF).  The work of K.~Y.\ is
supported in part by JSPS Research Fellowships for Young Scientists.


\begin{thebibliography}{99}

\bibitem{BFSS} T.~Banks, W.~Fischler, S.~H.~Shenker and L.~Susskind, 
``M theory as a matrix model: A conjecture,'' 
Phys.\ Rev.\ D {\bf 55} (1997) 5112 [arXiv:hep-th/9610043]. 

\bibitem{IKKT} N.~Ishibashi, H.~Kawai, Y.~Kitazawa and A.~Tsuchiya, 
``A large-N reduced model as superstring,'' 
Nucl.\ Phys.\ B {\bf 498} (1997) 467 [arXiv:hep-th/9612115]. 

\bibitem{DVV} R.~Dijkgraaf, E.~Verlinde and H.~Verlinde, 
``Matrix string theory,'' 
Nucl.\ Phys.\ B {\bf 500} (1997) 43 [arXiv:hep-th/9703030]. 

\bibitem{Witten} E.~Witten, 
``String theory dynamics in various dimensions,'' 
Nucl.\ Phys.\ B {\bf 443} (1995) [arXiv:hep-th/9503124]. 

\bibitem{dWHN} B.~de Wit, J.~Hoppe and H.~Nicolai, 
``On the quantum mechanics of supermembranes,'' 
Nucl.\ Phys.\ B {\bf 305} (1988) 545. 

\bibitem{BMN} D.~Berenstein, J.~M.~Maldacena and H.~Nastase, ``Strings
in flat space and pp waves from N = 4 super Yang Mills,'' JHEP {\bf
0204} (2002) 013 [arXiv:hep-th/0202021].

\bibitem{KG} J.~Kowalski-Glikman, ``Vacuum states in supersymmetric 
Kaluza-Klein theory,'' Phys.\ Lett.\ B {\bf 134} (1984) 194. 

\bibitem{DSR}
K.~Dasgupta, M.~M.~Sheikh-Jabbari and M.~Van Raamsdonk,
``Matrix perturbation theory for M-theory on a PP-wave,''
JHEP {\bf 0205} (2002) 056
[arXiv:hep-th/0205185]. 

\bibitem{SY}K.~Sugiyama and K.~Yoshida, ``Supermembrane on the pp-wave
background,'' Nucl.\ Phys.\ B {\bf 644} (2002) 113
[arXiv:hep-th/0206070]; ``BPS conditions of
supermembrane on the pp-wave,'' Phys.\ Lett.\ B {\bf 546} (2002) 143
[arXiv:hep-th/0206132]; 
N.~Nakayama, K.~Sugiyama and K.~Yoshida,
``Ground state of the supermembrane on a pp-wave,'' 
Phys.\ Rev.\ D {\bf 68} (2003) 026001, [arXiv:hep-th/0209081].

\bibitem{BSS}T.~Banks, N.~Seiberg and S.~H.~Shenker, 
``Branes from matrices,'' 
Nucl.\ Phys.\ B {\bf 490} (1997) 91 [arXiv:hep-th/9612157].

\bibitem{HS1} S.~Hyun and H.~Shin, ``Branes from matrix
theory in pp-wave background,'' Phys.\ Lett.\ B {\bf 543} (2002)
115, hep-th/0206090.

\bibitem{SY4}
K.~Sugiyama and K.~Yoshida, 
``Type IIA string and matrix string on pp-wave,'' 
Nucl.\ Phys.\ B {\bf 644} (2002) 128 [arXiv:hep-th/0208029]. 

\bibitem{HS2}
S.~Hyun and H.~Shin, 
`$`\mathcal{N}$ = (4,4) type IIA string theory on pp-wave background,'' 
JHEP {\bf 0210} (2002) 070 [arXiv:hep-th/0208074].

\bibitem{Sekino} 
Y.~Sekino and T.~Yoneya,
``From supermembrane to matrix string,'' 
Nucl.\ Phys.\ B {\bf 619} (2001) 22 [arXiv:hep-th/0108176].

\bibitem{Bonelli}
G.~Bonelli, 
``Matrix strings in pp-wave backgrounds from deformed super Yang-Mills 
theory,'' JHEP {\bf 0208} (2002) 022 [arXiv:hep-th/0205213].

\bibitem{Myers} R.~C.~Myers, ``Dielectric-branes,''
JHEP {\bf 9912} (1999) 022 [arXiv:hep-th/9910053].

\bibitem{DSR2}
K.~Dasgupta, M.~M.~Sheikh-Jabbari and M.~Van Raamsdonk, 
``Protected multiplets of M-theory on a plane wave,'' 
JHEP {\bf 0209} (2002) 021 [arXiv:hep-th/0207050].

\bibitem{KP}N.~Kim and J.~Plefka, 
``On the spectrum of pp-wave matrix theory,'' 
Nucl.\ Phys.\ B {\bf 643} (2002) 31 [arXiv:hep-th/0207034].

\bibitem{KP2}
N.~Kim and J.~H.~Park, ``Superalgebra for M-theory on a pp-wave,'' 
Phys.\ Rev.\ D {\bf 66} (2002) 106007 [arXiv:hep-th/0207061]. 

\bibitem{TM5}
J.~Maldacena, M.~M.~Sheikh-Jabbari and M.~Van Raamsdonk,
``Transverse fivebranes in matrix theory,''
JHEP {\bf 0301} (2003) 038
[arXiv:hep-th/0211139].

\bibitem{Bak} D.~Bak, ``Supersymmetric branes in PP wave
background,'' Phys.\ Rev.\ D {\bf 67} (2003) 045017, hep-th/0204033.

\bibitem{Park}
J.~H.~Park,
``Supersymmetric objects in the M-theory on a pp-wave,''
JHEP {\bf 0210} (2002) 032
[arXiv:hep-th/0208161]. 

\bibitem{sol}
 D.~Bak, S.~Kim and K.~Lee,
``All higher genus BPS membranes in the plane wave background,''
  arXiv:hep-th/0501202.

\bibitem{SY3} K.~Sugiyama and K.~Yoshida,
``Giant graviton and quantum stability in matrix model on PP-wave background,''
Phys.\ Rev.\ D {\bf 66} (2002) 085022
[arXiv:hep-th/0207190]. 

\bibitem{HSKY} H.~Shin and K.~Yoshida, 
``One-loop flatness of membrane fuzzy sphere interaction 
in plane-wave matrix model,'' Nucl.\ Phys.\ B {\bf 679} (2004) 99  
[arXiv:hep-th/0309258]. 

\bibitem{Huang}W.~H.~Huang, 
``Thermal instability of giant graviton in matrix model on pp-wave 
background,'' 
Phys.\ Rev.\ D {\bf 69} (2004) 067701 [arXiv:hep-th/0310212].

\bibitem{HSKY-thermal}
  H.~Shin and K.~Yoshida,
  ``Thermodynamics of fuzzy spheres in pp-wave matrix model,''
  Nucl.\ Phys.\ B {\bf 701} (2004) 380
  [arXiv:hep-th/0401014]; 
``Thermodynamic behavior of fuzzy membranes in PP-wave matrix model,''
  Phys.\ Lett.\ B {\bf 627} (2005) 188
  [arXiv:hep-th/0507029]. 

\bibitem{Furuuchi}
K.~Furuuchi, E.~Schreiber and G.~W.~Semenoff,
``Five-brane thermodynamics from the matrix model,''
arXiv:hep-th/0310286, \\ 
G.~W.~Semenoff, ``Matrix model thermodynamics,'' 
arXiv:hep-th/0405107; \\ 
S.~Hadizadeh, B.~Ramadanovic, G.~W.~Semenoff and D.~Young, 
``Free energy and phase transition of the matrix model on a plane-wave,'' 
Phys.\ Rev.\ D {\bf 71} (2005) 065016 [arXiv:hep-th/0409318].

\bibitem{HSKY-potential} 
H.~Shin and K.~Yoshida, 
``Membrane fuzzy sphere dynamics in plane-wave matrix model,'' 
Nucl.\ Phys.\ B {\bf 709} (2005) 69 [arXiv:hep-th/0409045]. 

\bibitem{KT}
D.~Kabat and W.~I.~Taylor, ``Spherical
membranes in matrix theory,'' Adv.\ Theor.\ Math.\ Phys.\ 
{\bf 2} (1998) 181 [arXiv:hep-th/9711078]. 

\bibitem{Kimura}
T.~Kimura and K.~Yoshida, 
``Spectrum of eleven-dimensional supergravity on a pp-wave background,''
Phys.\ Rev.\ D {\bf 68} (2003) 125007 [arXiv:hep-th/0307193]. 

\bibitem{wolfram} S. Wolfram, {\it the Mathematica book}, 4th edition
  (Cambridge, 1999).

\bibitem{LMW}
H.~K.~Lee, T.~McLoughlin and X.~k.~Wu, 
``Gauge / gravity duality for interactions of spherical membranes in 
11-dimensional pp-wave,'' 
Nucl.\ Phys.\ B {\bf 728} (2005) 1 [arXiv:hep-th/0409264].

\bibitem{future}
H.~Shin and K.~Yoshida, in preparation. 


\end{thebibliography}
\end{document}